**Comments by William M. Gray (Colorado State University) on the recently published paper in <u>Science</u> by Webster, *et al.*, titled "Changes in tropical cyclone number, duration, and intensity in a warming environment" (September 2005, Vol. 309, pp. 1844-1846, www.sciencemag.org).**

*ABSTRACT*

Recent US major landfalling hurricanes Katrina and Rita and last year's four U.S. landfalling major hurricanes have spawned an abundance of questions concerning the role that global warming might be playing in these events. This idea has been given added credence by the September 2005 *Science* paper of Webster, Holland, Currie and Chang (Vol. 304, pp. 1844-1846) showing that the global number of Category 4-5 hurricanes have increased in the last 15 years (1990-2004) in comparison with the prior 15-year period of 1975-1989. They report 171 Category 4-5 hurricanes in the earlier 15-year period vs. 269 (56% increase) in the later 15 year period. Global mean surface temperature in the later period has been about 0.3$^o$C higher than in the earlier period. The authors' imply that their measured rise in global Category 4-5 hurricanes is likely related to these higher global temperatures.

Having been involved with hurricane research and forecasting for nearly 50 years, I feel I have an obligation to offer comments on this paper's primary finding on the recent rise of global Category 4-5 hurricanes. I do not agree that global Category 4-5 tropical cyclone activity has been rising, except in the Atlantic over the last 11 years. The recent Atlantic upsurge has explanations other than global temperature rise.

## *DISCUSSION*

The near universal reference to this paper over the last few weeks by most major media outlets is helping to establish a false belief among the general public that global hurricane intensity has been rising and that global warming may be a contributing factor. I cannot accept the accuracy of the authors' measurements of global Category 4-5 hurricanes during 1975-1989 as indicated in their Table 1. This earlier 15 year global data set would not have been able to accurately delineate Category 4-5 hurricanes from Category 3 hurricanes or even at times from Category 1-2 hurricanes. It was just not possible to confidently distinguish the dividing line between maximum sustained surface winds above or below 130 mph (110 knots) in most global storm basins during the 1975-1989 period.

In the late 1970s I visited all the global tropical cyclone centers and observed their satellite capabilities and the training of their forecasters as part of a World Meteorological Organization (WMO) tropical cyclone survey trip that I was commissioned to make. The satellite tools and forecaster training in the tropical cyclone regions of the Indian Ocean and Southern Hemisphere during the 1975-1989 period was not adequate for the task of objectively distinguishing Category 4-5 hurricanes from Category 3 hurricanes or to always be able to confidentially distinguish Category 4-5 hurricanes from Category 1-2 hurricanes. Table 1 of the Webster et al. paper indicates that there were 32 Indian Ocean and South Pacific Category 4-5 tropical cyclones in 1975-89 and 79 (247 percent more) during the 15-year period of 1990-2004. Such large increases are not reasonable given our lack of confidence in the Category 4-5 measurement techniques and the fact that the frequencies of the weaker cyclones in these basins did not show much difference.

This paper also presents data which shows that there has been no general increase in the number of global hurricanes and tropical storms over the last 35 years during which global sea-surface temperatures have been rising. I concur with this measurement. It agrees with the recent research by my colleague, Phil Klotzbach, who has made similar tabulations.

### *DETERMINATION OF TROPICAL CYCLONE MAXIMUM WIND SPEEDS*

There always has been, and there probably always will be, problems in assigning a representative maximum sustained surface wind to a hurricane. As technology advances and the methods of determining a tropical cyclone's maximum sustained surface winds change, different values of maximum winds will often be assigned to hurricanes than the values that would have been assigned in previous years.

With the availability of new aircraft deployed inertial dropwindsondes and the new step-frequency surface wind measurement instruments in the Atlantic, it is being found that Atlantic hurricanes and some Northeast Pacific hurricanes that were flown have sustained surface winds that are often stronger than would have been estimated from wind values extrapolated from aircraft altitude. Due to these recent and continuing changes in measurement techniques (Franklin et al. 2003), Saffir-Simpson category numbers in the Atlantic have and likely will continue to creep upward. These changes will likely be translated to other global tropical basins.

### *CHANGE IN INTENSITY MEASUREMENT TECHNIQUES IN THE NORTHWEST (NW) PACIFIC*

The Northwest Pacific basin is the most active of all tropical cyclone basins. It had aircraft reconnaissance center fixes during the period 1945-1986 but has not had aircraft reconnaissance since. The satellite has been the only tool to track NW Pacific typhoons since 1987.

There was an anomaly in the measurement of typhoon intensity in the 14-year period of 1973-1986 when the Atkinson-Holliday (1977) technique for typhoon maximum wind ($V_{max}$) and minimum sea-level pressure (MSLP) was used. The Atkinson-Holliday (AH) technique is known to have significantly underestimated the maximum winds of typhoons in comparison with their central pressures. This interpretation has been supported by a combination of comparative satellite-aircraft data from the Atlantic; by pre-1973 NW Pacific aircraft-measured wind-pressure, and by the pure satellite measurement since 1987. This topic has been extensively reviewed by Knaff and Zehr (2005). Table 1 shows the official average of the annual number of super typhoons in the West Pacific (equivalent to the number of category 3-4-5 or major hurricanes of the Atlantic). Note that between 1950-1972 and over the last 18 years (1987-2004), the number of super-typhoons has averaged about five per year while during the Atkinson-Holliday period of 1973-1986 it was less than half this number. Yet weaker storm frequency during the 1973-1986 period was about the same as in the earlier and later periods. If we disregard this anomalous 1973-1986 period and compare annual frequency of super-typhoon activity between 1950-1972 versus 1987-2004, we see little difference despite the recent global warming trend.

*Table 1. Comparison of the annual average of super-typhoon activity in three multi-decadal periods in the western North Pacific. The middle period (1973-1986) used the Atkinson-Holliday (1977) intensity scheme. Reported maximum wind values were too low.*

| Years | Annual Average Number of Super-Typhoons | Basin July-Sept SST ($^o$C) 10-25$^o$N; 120-160$^o$E |
|---|---|---|
| 1950-1972 | 5.3 | 28.93 |
| 1973-1986 (AH) | 2.3 | 28.92 |
| 1987-2004 | 4.9 | 29.22 |

Who would believe that the annual super typhoon activity of the 14 year period 1973-86 (years AH technique was applied) would be only 44 percent of the annual super typhoon activity of the prior 23-year period or the current 18-year period? This period of suppressed super-typhoon frequency during 1973-86 closely corresponded with the first 15-year period of Webster et al.'s 1975-89 Category 4-5 data.

## *VARIATION IN MAJOR HURRICANE NUMBERS DURING THE LAST TWO DECADES OF GLOBAL WARMING*

As tropical cyclone maximum wind ($V_{max}$) observational techniques are frequently not adequate to distinguish between Category 4-5 and Category 3 hurricanes, it might be more representative to observe the increase of major hurricanes (Category 3-4-5). There has been US-Japanese satellite coverage of the north Pacific during the last 20 years, and both satellite and aircraft reconnaissance data have been available in the Atlantic. The biggest rise in global surface air temperature occurred during the last 10 years. The NOAA-NCEP reanalysis of global mean temperature differences between the last two 10-year periods show that the last 10 years (1995-2004) of global surface temperature have been about 0.4°C warmer than the earlier 10-year period of 1985-1994. If there was an influence of global warming on major hurricane activity, one would expect to see this increase represented by greater numbers of global major hurricanes during the last 10 years in comparison with the earlier 10-year period.

Table 2 shows the number of measured major hurricanes (Cat. 3-4-5) around the globe (excluding the Atlantic). Note that there has been no apparent difference in reported major (Cat. 3-4-5) hurricanes between these two 10-year periods despite the globe being about 0.4°C warmer in the recent period.

*Table 2. Comparison of observed major (Cat. 3-4-5) tropical cyclones in all global basins (except the Atlantic) in the two most recent 10-year periods of 1985-94 and 1995-2004.*

|  | **1985-1994 *(10 Years)*** | **1995-2004 *(10 Years)*** |
|---|---|---|
| North & South Indian Ocean | 45 | 50 |
| South Pacific & Australia | 44 | 41 |
| NW Pacific | 88 | 87 |
| Northeast Pacific | 41 | 40 |
| GLOBE TOTAL (excluding Atlantic) | 218 | 218 |

By contrast, the Atlantic has seen a very large increase in major hurricanes during the last 10-year period in comparison to the previous 10-year period (*38* between 1995-2004 vs. *14* during 1985-1994). The large increase in Atlantic major hurricanes during the last 10 years is primarily a result of the multi-decadal increase in the Atlantic Ocean thermohaline circulation (THC) and not due to global temperature increase. Changes in salinity are believed to be the driving mechanism. These multi-decadal changes have also been termed the Atlantic Multi-Decadal Oscillation (AMO). Even when the large increase in Atlantic major hurricane activity is added to the non-Atlantic global total of major hurricanes, there is no significant global difference (232 vs. 256) in the numbers of major hurricanes between these two most recent 10-year periods.

***COMPARISON OF PACIFIC CATEGORY 4-5 TROPICAL CYCLONE ACTIVITY DURING THE LAST TWO 10-YEAR PERIODS***

The most reliable comparison of Category 4-5 hurricanes that can likely be made is to compare the last ten years (1995-2004) with the prior ten years (1985-1994) for the storm areas monitored by the US and Japan. The two North Pacific basins do not indicate that the number of hurricanes of Category 4-5 intensity have increased in the last 10 years when global surface temperature have risen (Table 3).

Table 3. Comparison of the number of Category 4-5 hurricanes in the North Pacific during the last two 10-year periods.

|  | 1985-1994 (10 Years) | 1995-2004 (10 Years) |
|---|---|---|
| NE PACIFIC | 31 | 30 |
| NW PACIFIC | 70 | 65 |
| *TOTAL* | *101* | *95* |

## COMPARISON OF ATLANTIC HURRICANE ACTIVITY BETWEEN THE LAST 15-YEAR ACTIVE PERIOD (1990-2004) WITH AN EARLIER ACTIVE 15-YEAR PERIOD (1950-1964)

There have been past hurricane periods in the Atlantic which have had just as many major hurricanes and Category 4-5 hurricanes as in recent years. A comparison of the last 15 years of hurricane activity with an earlier 15-year period from 1950-64 shows no significant difference in major hurricanes or Category 4-5 hurricanes (Table 4) even though the global surface temperatures were colder and there was a general global cooling during 1950-64 as compared with global warming during 1990-2004. The maximum sustained winds from 1950-1964 have been adjusted downward using the Landsea (1993) adjustment factor.

The number of weak tropical storms rose by over 50 percent during this later 15 year period, however. This is a reflection of the availability of the satellite in the later period. It would not have been possible that a hurricane, particularly a major hurricane, escaped detection in the earlier period. But many weaker systems far out in the Atlantic undoubtedly went undetected before satellite observations.

*Table 4. Comparison of Atlantic tropical cyclones of various intensities between 1950-1964 and the recent 15 year period of 1990-2004.*

|  | Cat. 4-5 | Cat. 3 | Net IH | Net H | Cat. 1-2 | TS | NS | July-August SST 10-25°N; 30-70°W |
|---|---|---|---|---|---|---|---|---|
| 1950-64 (15 yrs) | 24 | 23 | 47 | 98 | 51 | 50 | 148 | 25.69 |
| 1990-04 (15 yrs) | 25 | 18 | 43 | 100 | 57 | 78 | 178 | 26.11 |
| 1990-04 *minus* 1950-64 | +1 | -5 | -4 | +2 | +6 | +28 | +30 | +0.42 |
| Percent Increase | +4% | -22% | -9% | +2% | +12% | +56% | +18% | --- |

## *SUMMARY*

Despite what many in the atmospheric modeling and forecast communities may believe, there is no physical basis for assuming that global tropical cyclone intensity or frequency is necessarily related to global mean surface temperature changes of less than ±0.5°C. As the ocean surface warms, so does the global upper air temperature to maintain conditionally unstable lapse-rates and global rainfall rates at their required values. Seasonal and monthly variations of SST within individual storm basins show only very low correlations with monthly, seasonal, and yearly variations of hurricane activity. These correlations are typically of the order of about 0.3, explaining only about 10 percent of the variance. Other factors such as tropospheric vertical wind shear, surface pressure, low level vorticity, mid-level moisture, etc. play more dominant roles in explaining hurricane variability on shorter time scales. Although there has been a general global warming over the last 30 years and particularly over the last 10 years, the SST increases in the individual tropical cyclone basins have been smaller (about half) and, according to the observations, have not brought about any significant increases in global major tropical cyclones except for the Atlantic which as discussed, has multidecadal oscillations driven primarily by changes in salinity. No credible observational evidence is available or likely will be available in the next few decades which will be able to directly associate global temperature change to changes in global Category 4-5 hurricane frequency and intensity.

*Acknowledgement*: I would like to acknowledge beneficial discussions on this topic with John Knaff and Philip Klotzbach.